\documentclass[12pt]{article}

\usepackage{latexsym,amsmath,amssymb,epsfig}

\def\be{\begin{equation}}
\def\ee{\end{equation}}

\def\bea{\begin{eqnarray}}
\def\eea{\end{eqnarray}}
\def\nn{\nonumber \\}

\def\ie{{\it i.e.\/}}

\def\uneq{{\, \not \! = \,}}

\def\eps{{\epsilon}}

\def\bpsi{\bar\psi}
\def\del{\delta}

\def\hc{{\rm h.c.}}

\newcommand{\dif}{{\mathrm{d}}}

\def\d{\partial}
\def\dsl{{\not \! \partial}}

\newcommand{\refeq}[1]{\mbox{(\ref{#1})}}
\newcommand{\ltsim}{\lower3pt\hbox{$\, \buildrel < \over \sim \, $}}
\newcommand{\gtsim}{\lower3pt\hbox{$\, \buildrel > \over \sim \, $}}

\makeatletter 
\def\secteqno{\@addtoreset{equation}{section}%
\def\theequation{\thesection.\arabic{equation}}}

\begin{document}

\secteqno
\baselineskip=.58cm


\renewcommand{\thefootnote}{\fnsymbol{footnote}}

{\hfill \parbox{5cm}{ {\footnotesize
	CAFPE-16/03, UG-FT-146/03 \\
        DFPD 03/TH/07 \\
	IPPP/03/04, DCPT/03/08 } \\ 
        hep-th/0302023 }} 

\bigskip\bigskip\bigskip
\begin{center}
{\bf \Large Bulk fields with general brane kinetic terms}

\bigskip\bigskip

F. del Aguila$^{a}$,
M. P\'erez-Victoria$^{a,b}$ and
J. Santiago$^{c}$

\bigskip\bigskip

{\it
\noindent
\mbox{}$^a$ Departamento de F\'{\i}sica Te\'orica y del Cosmos and
Centro Andaluz de F\'{\i}sica de Part\'{\i}culas Elementales (CAFPE),
Universidad de Granada, E-18071 Granada, Spain

\noindent
\mbox{}$^b$ Dipartimento di Fisica ``G. Galilei'', Universit\`a di
Padova and INFN, Sezione di Padova, Via Marzolo 8, I-35131 Padua,
Italy 

\noindent
\mbox{}$^c$ Institute for Particle Physics Phenomenology, University
of Durham, South Road, Durham DH1 3LE, UK

}

\vspace{2cm}
{\bf Abstract}
\vspace{1cm}

\parbox{12.5cm}{
We analyse the effect of general brane kinetic terms for bulk scalars,
fermions and gauge bosons in theories with extra dimensions, with and
without supersymmetry. We find in particular a singular behaviour when
these terms contain derivatives orthogonal to the 
brane. This is brought about by $\delta(0)$ divergences arising at
second and higher order in perturbation theory. We argue that
this behaviour can be smoothed down by classical renormalization.
}

\end{center}

\renewcommand{\thefootnote}{\arabic{footnote}}

\vspace{3cm}

\section{Introduction}

Field theoretical models with extra dimensions typically include one
or more lower dimensional defects on which some fields can be
localized. These defects (``branes'' from now on) can have different
microscopic realizations: D-branes, domain walls, cosmic strings,
orbifold fixed points, \ldots. 
In general, the fields propagating in
all the dimensions (bulk fields) couple 
to fields propagating on the
branes~\cite{Horava:1995qa,Mirabelli:1997aj}. Moreover, 
it is possible that the action includes terms localized on the branes
involving only bulk fields (``brane terms'', for short). 
They may have important consequences. 
A crucial observation is that brane terms are
induced by radiative corrections. This is allowed as Poincar\'e
invariance is broken by the branes. Quantum divergences localized on
the branes can appear in the correlators of
bulk fields when the bulk fields are coupled to fields living
on them~\cite{Dvali:2000hr}. Furthermore, in orbifolds this can
occur even in the absence of brane couplings at tree
level~\cite{Georgi:2000ks}. In any case, it is
clear that the presence of divergent radiative 
corrections localized on the branes requires the introduction of
divergent brane counterterms. Therefore, the corresponding brane
couplings run with the renormalization scale, and cannot be set to
zero at all scales. Furthermore, this indicates that they should be
considered free parameters of the theory and be included at
tree level (see~\cite{Carena:2002me} for a detailed argument). Of
course, since
interacting theories in higher dimensions are nonrenormalizable, 
these considerations are only meaningful in the framework of
effective field theory, and it is important to take into account
possible suppressions of the brane couplings by inverse powers of the
cutoff scale $\Lambda$. 

Brane terms are always suppressed by $1/\Lambda^n$, $n\geq1$,
with respect to the analogous bulk terms. Therefore one would
expect that their effect on observables should be a small perturbation
of the corresponding results without brane terms. In this article we
show that, contrary to this expectation, certain brane operators
modify severely some observables---like Kaluza-Klein (KK)
masses---even when their coefficients are 
very small. This breakdown of perturbation theory has to do with the
interplay between the thin brane limit and the limit of large
$\Lambda$, in which the coefficients of the brane operators approach
zero. Unless other\-wise stated, we will always
consider infinitely thin, delta-function branes, although we will
often make use of thick branes as intermediate regulators.
On the other hand, the underlying fundamental
theory is usually assumed to be smooth in its parameters, so that all
its low energy predictions should be perturbatively calculable using
some effective theory. From this point of view, the relevant question
is how to incorporate in a low energy effective description a
mechanism to make the expansion in $\Lambda$ well defined. Here we
sketch a possible answer in the spirit of perturbative renormalization
theory. 

We shall concentrate on brane kinetic terms. Kinetic terms localized
on a brane were first invoked 
by Dvali, Gabadadze and Porrati~\cite{Dvali:2000hr} to recover 4D 
gravitational behaviour at short distances in infinite extra
dimensions. The same 
mechanism was subsequently applied in~\cite{Dvali:2000rx} to gauge
theories. In~\cite{Cheng:2002iz}, the one loop 
radiative corrections to brane terms and their first order impact on
KK spectra were calculated in different 
theories. In~\cite{vonGersdorff:2002as}, a detailed computation of
infinite and finite radiative 
corrections to bulk and brane kinetic terms of gauge fields (with and
without Hosotani mechanism) was performed.  
The phenomenology of gauge fields in compact dimensions with
(tree-level) brane kinetic terms has been discussed recently
in~\cite{Carena:2002me} (flat space)
and~\cite{Carena:2002dz,Davoudiasl:2002ua} (warped 
space). A particular fermionic case has been addressed
in~\cite{delAguila:2002mt}. Finally, brane kinetic terms in
supersymmetric models have been studied
in~\cite{Hebecker:2001ke,Cheng:2002ab,Kyae:2002fk,Chaichian:2002uy}.
In all these works, brane kinetic terms 
are found to have a significant impact on model building and
phenomenology. And, as observed above, they must be generically
included for consistency. 

Here we analyse at the classical level the effect of all the possible
$\mathcal{O}(1/\Lambda)$ brane kinetic terms for scalar, fermion and
gauge 
fields. In particular we examine brane terms containing derivatives 
with respect to the coordinates orthogonal to the brane. These terms
have not been studied in detail before\footnote{Except
in~\cite{Lewandowski:2001qp}, where the singular terms were
``massaged'' via a 
classical renormalization. We compare this approach with ours in
Section~\ref{renormalization}.}. We find that they give rise to
solutions to the 
equations of motion which 
exhibit a rather singular behaviour: they decouple
from the brane and become insensitive to the magnitude of all brane
couplings; furthermore, physical observables for arbitrarily small
coefficients of these terms do not approach the ones for 
vanishing coefficients. This is related to the appearance of
$\delta(0)$ singularities at second and higher orders in the
$1/\Lambda$ expansion. These singularities are a common feature of
field theoretical models with infinitely thin
branes~\cite{Horava:1996ma,Mirabelli:1997aj}. We also study brane
kinetic terms in supersymmetric theories and find that supersymmetry
requires the presence of higher order kinetic terms in the on-shell
action. These terms improve but do not cure the singular
behaviour of the classical fields. More generally, any
effective theory will contain higher order terms which may
significantly change the solutions derived from the first order
Lagrangian. One can then hope that these terms
be such that the exact solutions are smooth in all
the parameters of the theory. We will discuss this possibility below. 
Although most of our considerations apply to general
field theories in extra dimensions, for definiteness
we study here theories defined on a flat 5D space with
the 5th dimension compatified on an $S^1/Z_2$ orbifold. Moreover, in
order to keep the discussion and calculations as simple as possible,
we neglect masses and consider kinetic terms localized on 
only one of the fixed points (even if kinetic terms on both branes are
natural and lead to interesting phenomenology~\cite{Carena:2002me}). 

The structure of the paper is as follows. In Section~\ref{exact} we
perform the exact KK reduction of the free action for particles of
spin $\leq 1$ in the presence of general brane kinetic terms.
In Section~\ref{susy} we study the supersymmetric case. In
Section~\ref{singular} we examine in detail an example elucidating the
role of thin brane singularities. In Section~\ref{renormalization} we
argue that classical renormalization makes the theory smooth and
perturbation theory well defined. This is also interpreted as the
restriction to a subclass of effective theories with particular
reductions of couplings. We summarize and discuss our results in
Section~\ref{conclusions}. Finally, we collect in
Appendix~\ref{firstorderappendix} the general first order results, and
give an example of renormalization in Appendix~\ref{renappendix}. 


\section{Exact KK reduction}
\label{exact}

In this section we study the exact (\ie, nonperturbative) KK reduction
of the most general free action with leading order brane terms for
massless scalars, gauge bosons and fermions propagating in $M_4\,
\times \, S^1/Z_2$. We consider fields which are functions of the
coordinates $x^\mu$, $\mu=0,1,2,3$ of $M_4$ and $y\in (-\pi
R,\pi R]$, of $S^1$, and which are eigenstates of $Z_2$, \ie,
even or odd functions of $y$. For simplicity, we only include
brane terms at the fixed point $y=0$. 

Some of the operators we consider are products of two odd
factors times a delta function. These terms are sometimes argued to
vanish on parity grounds. However, these arguments are no longer valid
when the (derivatives of) wave functions appearing in these terms
develop discontinuities across the
branes~\cite{Meissner:2002dg}. Deriving and 
solving the equations of motion when the action contains this sort of
terms is tricky~\cite{Bagger:2001qi}, and the same applies to 
brane terms containing derivatives with respect to $y$. 

In order to
treat all these cases rigorously we regularize the branes using
several (smooth) representations of the delta function, of width~$\sim
\eps$. We perform all the 
calculations with these resolved branes, and only take the thin brane
limit $\eps \rightarrow 0$ at the very end. Furthermore, we employ
different representations in order to check that our results are
regularization independent. In some cases we have resorted to
numerical methods to 
solve the relevant differential equations. Still, it is possible to
see that the numerical solutions converge rapidly in 
the thin brane limit to certain functions, which we take as our
analytical solutions. The reader can find an explicit analytical
calculation in the simple case discussed in Section~\ref{singular}.

\subsection{Scalars}

The most general action for a massless complex five-dimensional
scalar with leading order brane kinetic terms is\footnote{More
generally we 
could consider a term $\delta_0(\frac{b}{2} \phi^\dagger
\partial_y^2 \phi + \frac{b^\ast}{2} \partial_y^2 \phi^\dagger \phi)$,
but we take $b$ real for simplicity.}
\begin{align}
S=&\; \int \dif^4 x \; \int_{-\pi R}^{\pi R} \dif y\; \Big\{
\big(1+a \delta_0 \big) \partial_\mu \phi^\dagger \partial^\mu \phi 
- \partial_y \phi^\dagger \partial_y \phi 
\nonumber \\
&\; \phantom{\int \mathrm{d}y\;} +
\delta_0\big[\frac{b}{2}(\phi^\dagger \partial_y^2 \phi + \partial_y^2
  \phi^\dagger \phi)-c \partial_y \phi^\dagger \partial_y \phi 
\big] \Big\} \, , \label{scalaraction}
\end{align}
where $\delta_0\equiv \delta(y)$. The dimensionful parameters $a$, $b$
and $c$ are expected to be of order $1/\Lambda$.
In order to perform the KK reduction we expand
\begin{equation}
\phi(x,y)=\sum_{n=0}^\infty \frac{f_n(y)}{\sqrt{2\pi R}}
\phi^{(n)}(x) \, , \label{scalarKKexpansion}
\end{equation}
where $\{f_n\}$ is some basis of functions on $(-\pi R,\pi R]$ with the
same parity character as $\phi$. Introducing~\refeq{scalarKKexpansion}
in the action we find
\begin{align}
S=& \;
\int \dif^4 x \; \sum_{mn} \Big\{ \int_{-\pi R}^{\pi R} \dif y\; (1+a
\delta_0)\frac{f_m f_n}{2\pi R}  \partial_\mu
\phi^{(m)\dagger} \partial^\mu \phi^{(n)}
\nonumber \\
&\;
\phantom{\sum_{mn} \Big\{}+
\int \mathrm{d}y\; \frac{f_m 
\mathcal{O}
f_n}{2\pi R}  
\phi^{(m)\dagger}  
\phi^{(n)}
\Big\} \, ,
\end{align}
where we have defined
\begin{equation}
\mathcal{O}\equiv
[1 +(b+c)\delta_0]\partial_y^2 +(b+c)\delta^\prime_0 \partial_y
+\frac{b}{2} \delta^{\prime\prime}_0 \, . 
\end{equation}
In order to diagonalize the quadratic terms, we take $\{f_n\}$ as the
set of eigenfunctions of the following generalised eigenvalue problem:
\begin{equation}
\mathcal{O} f_n=-m_n^2 \big(1+a \delta_0\big) f_n. \label{eigeneq}
\end{equation}
The fact that the operator $\mathcal{O}$ is hermitian and that the
eigenvalues of~\refeq{eigeneq} are nondegenerate, ensures the
orthogonality of eigenfunctions with respect to the scalar product
\begin{equation}
<f,g> = \frac{1}{2\pi R} \int_{-\pi R}^{\pi R} \dif y \big (1+a
\del_0 \big) f(y) g(y) \, .  \label{scalarprod}
\end{equation}
Then, if we also impose, when possible, the normalization
\begin{equation}
<f_n,f_n> = 1 \, , \label{normalizationcond}
\end{equation}
the free action reduces to
\begin{equation}
S= \;
\int \dif^4 x \; \sum_n \, \Big\{ \partial_\mu \phi^{(n)\dagger}
\partial^\mu \phi^{(n)} - m_n^2 \phi^{(n)\dagger} \phi^{(n)}
\Big\} \, .
\end{equation}
The generalised eigenvalue equation~\refeq{eigeneq} is to be solved
with the appropriate orbifold constraints. 

When $b<0$ or $c<0$, it turns out that, for finite (small enough)
$\eps$, the eigenfunctions diverge at some point near 
$y=0$. For this reason, we must require $b\geq0$, $c\geq0$ to have a
well defined theory.

Let us now describe the exact solutions in the thin brane limit. 
For odd scalars, the wave functions and KK masses are insensitive to
the brane terms. The wave functions are
\begin{equation}
f_n^\mathrm{odd}(y)=\sqrt{2} \sin(m_n y),
\end{equation}
and the masses, $m_n=n/R$, with $n$ a positive integer\footnote{The
derivatives of the wave functions, however, have discontinuous values
at $y=0$ in the limit $\eps \rightarrow 0$, whenever $c \uneq 0$:
$f_n^\prime(0)= 0 \uneq f_n^\prime(0^+)$. This behaviour at the brane
is the same as the one for the even component of a fermion, to be
described below.}.

For even scalars, the wave functions and masses do not depend on the
(positive) coefficient $c$ when $\eps\rightarrow 0$. We
distinguish two cases:   
\begin{itemize}

\item $b=0$. In this case, the solution coincides with the one
described in~\cite{Carena:2002me}. There is a constant massless mode,
which depends on the brane terms only through its normalization:
\begin{equation}
f_0=\frac{1}{\sqrt{1+\frac{a}{2\pi R}}} \, .
\end{equation}
The massive modes are given by
\begin{equation}
f_n(y)=A_n [\cos(m_n y)-\frac{a m_n}{2} \sin(m_n |y|)], 
\end{equation}
where $A_n$ is a normalization constant and
the masses are the solutions of the equation
\begin{equation}
\tan(\pi R m_n)+\frac{a}{2}m_n=0.
\label{Carenatwobranes:eigen}
\end{equation}
We observe that if $-2\pi R \leq a<0$, there is a tachyonic
mode, with squared mass approaching $-4/a^2$ for small $|a|/(2\pi
R)$. On the other hand, the norm squared defined by
the scalar product~\refeq{scalarprod}, $||f||^2=\mbox{$<f,f>$}$, is
indefinite. This can be problematic, since zero norm modes have
vanishing kinetic term in the reduced 4D action (they cannot be
normalized as in~\refeq{normalizationcond}), while modes with negative
norm squared have the wrong sign in the reduced kinetic term (they are
ghosts). 
We find that if $a=-2 \pi R$ the zero mode has zero norm,
and if $a<-2\pi R$ it has negative norm squared (the massive modes
have always positive norm). 
Therefore, we see that for negative values of $a$ the theory is either
pathological or unstable.

\item $b > 0$. In this case the solution is independent of all
brane terms, including $b$ itself (as long as it is strictly
positive). We find 
\begin{equation}
f_n(y)=A_n \sin(m_n |y|), 
\end{equation}
with $m_n=\frac{n+1/2}{R}$ and $n$ integer. In particular, there is no
zero mode. 

\end{itemize}
We see that the KK spectrum and wave functions of even scalars change
dramatically when an arbitrarily small $b$ is turned on. Note also
that the wave functions vanish at $y=0$ when
$b>0$.

\subsection{Gauge Bosons}

The Lagrangian for gauge bosons admits brane kinetic terms analogous
to the ones for scalars, but with the restrictions of gauge
invariance. Since we are only discussing the quadratic action, we can
consider an abelian theory without loss of 
generality.
For a gauge boson $A_M$ with general
$\mathcal{O}(1/\Lambda)$ kinetic terms on the first brane,
\begin{equation}
S = \int \dif^4 x \int \dif y \, \left\{-\frac{1}{4}
(1+a\del_0)  F_{\mu\nu}F^{\mu\nu} -\frac{1}{2} (1+c\del_0)
F_{5\nu}F^{5\nu} \right\} \, . \label{gaugeLag}
\end{equation}
The bulk and brane parts respect 5D and 4D Lorentz symmetry,
respectively. The orbifold projection forces $A_\mu$ and $A_5$ to have
opposite $Z_2$ parity. The whole Lagrangian is symmetric under the gauge
transformations $A_M \rightarrow A_M+\d_M \theta$, where the gauge
parameter $\theta$ has the same parity as $A_\mu$. Using a gauge
$A_5=0$ or $\d_M A^M=0$ it is easy to see that the equations of
motion are the same as the ones for scalars without $b$ terms. At the
level of the action, these gauges can be imposed adding the
corresponding gauge fixing terms. To perform the KK decomposition,
however, we choose to work with 
the gauge invariant action. We expand
\begin{eqnarray}
&&A_\mu(x,y)=\sum_{n=0}^\infty \frac{f_n(y)}{\sqrt{2\pi R}}
A_\mu^{(n)}(x) \, , \\
&&A_5(x,y)= \sum_{n=0}^\infty
\frac{\d_y f_n(y)}{m_n \sqrt{2\pi R}}
A_5^{(n)}(x) \,  ,  
\end{eqnarray}
with $A_5^{(0)}=0$ if $m_0=0$, and require
\begin{equation}
\mathcal{O} f_n = -m_n^2 (1+a\del_0) f_n \, ,
\end{equation}
with
\begin{equation}
\mathcal{O} \equiv (1+c\del_0) \d_y^2 + c \del^\prime_0 \d_y \, .
\end{equation}
Thanks to the form of the expansion of $A_5$ and the hermiticity of
the differential operator $\mathcal{O}$, all the terms get
diagonalized. To 
have canonical 4D kinetic terms we also impose the normalization
condition~\refeq{normalizationcond}.
The eigenfunctions and eigenmasses can be read directly from the
results for (even or odd) scalars with $b=0$. Again, there is no
well defined solution for $c<0$, so we require $c\geq 0$. And
for $a<0$ we have either tachyons or ghosts.

At the end we are left with the following 4D Lagrangian:
\begin{equation}
\mathcal{L} = \sum_{n=0}^{\infty} \, \left\{-\frac{1}{4} F_{\mu\nu}^{(n)}
F^{\mu\nu(n)} + \frac{1}{2} \d_\mu A_5^{(n)} \d^\mu A_5^{(n)}
+ \frac{1}{2} m_n^2 A_\mu^{(n)} A^{(n)\mu} + m_n A_5^{(n)} \d_\mu
A^{(n)\mu} \, \right\} \, .
\label{simpleaction}
\end{equation}
This Lagrangian is invariant under an infinite set of 4D
local transformations:
\begin{eqnarray}
&& A_\mu^{(n)} \rightarrow A_\mu^{(n)} + \d_\mu \theta^{(n)} \, , \\ 
&& A_5^{(n)} \rightarrow A_5^{(n)} + m_n \theta^{(n)} \, , 
\end{eqnarray}
which is nothing but the KK decomposition of 5D gauge transformations:
\begin{equation}
\theta(x,y) = \sum_{n=0}^\infty \frac{f_n(y)}{\sqrt{2\pi R}}\,
\theta^{(n)}(x) \, .
\end{equation}
The scalars $A_5^{(n)}$ can be interpreted as Goldstone bosons of the
spontaneously broken gauge transformations $\theta^{(n)}$ (those with
$f_n(y) \uneq \mbox{constant}$). They can be decoupled from the 4D
vector bosons $A_\mu^{(n)}$ using an $R_\xi$ gauge. Indeed, adding to
\refeq{simpleaction} the gauge fixing terms 
\begin{equation}
-\frac{1}{2\xi} \sum_{n=0}^\infty (\d_\mu A^{(n)\mu} + \xi m_n
A_5^{(n)})^2 \, , \label{KKgaugefixing}
\end{equation}
we find
\begin{align}
\mathcal{L}^\prime = \sum_{n=0}^{\infty} \, \Big\{&-\frac{1}{4}
F_{\mu\nu}^{(n)} F^{\mu\nu(n)} - \frac{1}{2\xi} (\d_\mu A^{(n)\mu})^2
+\frac{1}{2} m_n^2 A_\mu^{(n)} A^{(n)\mu} + \frac{1}{2} \d_\mu A_5^{(n)}
\d^\mu A_5^{(n)} \nn
& - \xi m_n^2 (A_5^{(n)})^2  \, \Big\} \,
. \label{gaugefixedKK} 
\end{align}
In the Feynman-'t Hooft gauge, $\xi=1$, the scalar fields $A_5^{(n)}$
have mass $m_n$, whereas in the unitary gauge, $\xi \rightarrow
\infty$, these fields get infinite masses and decouple from the
spectrum.

\subsection{Fermions}

Let us consider now the most general kinetic action for 
a five-dimensional bulk fermion $\psi=\psi_L+\psi_R$, with $\gamma^5
\psi_{L,R}=\mp \psi_{L,R}$, and leading order brane contributions, 
\begin{align}
S=& \int \dif^4 x \int_{-\pi R}^{\pi R} \mathrm{d}y\;
\Big\{
\big(1+a^L\delta_0 \big)
\bar{\psi}_L \mathrm{i} \not \! \partial \psi_L 
+ 
\big(1+a^R\delta_0 \big) 
\bar{\psi}_R \mathrm{i} \not \! \partial \psi_R
\nonumber \\
&
\phantom{\int_{-\pi R}^{\pi R} \mathrm{d}y}
-\big(1+\frac{b}{2}\delta_0 \big) \bar{\psi}_L \partial_y \psi_R
- \frac{b}{2} \delta_0 \big( \partial_y \bar{\psi}_R\big) \psi_L
\nonumber \\
&
\phantom{\int_{-\pi R}^{\pi R} \mathrm{d}y}
+\big(1+\frac{c}{2} \delta_0 \big) \bar{\psi}_R \partial_y \psi_L
+ \frac{c}{2} \delta_0 
\big( \partial_y \bar{\psi}_L\big) \psi_R
\Big\}. \label{fermion:lagrangian:5D}
\end{align}
We have considered again real $b$ and $c$.
The two chiralities have opposite parity. We take $\psi_L$ even for
definiteness. The KK expansion of the fermions reads
\begin{equation}
\psi_{L,R}(x,y)=\sum_n \frac{f_n^{L,R}(y)}{\sqrt{2 \pi R}}
\psi^{(n)}_{L,R}(x) \, .
\end{equation}
In order to diagonalize the 4D kinetic and mass terms, we take the
wave functions $f_n^{L,R}$ to be the eigenfunctions of the following 
generalised eigenvalue problem:
\begin{align}
\mathcal{O}_1 f_n^R =& m_n(1+a^L \delta_0) f_n^L \, ,
\label{eigenval:diff:1} 
\\
\mathcal{O}_2 f_n^L =&- m_n(1+a^R \delta_0)f_n^R \, ,
\label{eigenval:diff:2}  
\end{align}
where
\begin{align}
\mathcal{O}_1 =&  [1+\frac{1}{2}(b+c)\delta_0]\partial_y+\frac{c}{2}
\delta^\prime_0 \, , \\ 
\mathcal{O}_2 =&  [1+\frac{1}{2}(b+c)\delta_0]\partial_y+\frac{b}{2}
\delta^\prime_0 \, . 
\end{align}
Because $\mathcal{O}_1^\dagger = - \mathcal{O}_2$, the functions
$f_n^{L,R}$ are 
orthogonal with respect to the appropriate scalar products,
\begin{equation}
<f,g>_{L,R} = \frac{1}{2\pi R} \int_{-\pi R}^{\pi R} \mathrm{d}y\;
\big(1+a^{L,R} \delta_0 \big) \, f(y) g(y) \, .
\end{equation}
We also impose the normalization conditions
\begin{equation}
<f_n^L,f_n^L>_L= <f_n^R,f_n^R>_R=1 \, , 
\label{orthonormality} 
\end{equation}
to render the 4D kinetic terms canonically normalized.

As in previous cases, we must require $b,c \geq 0$ to have
well defined solutions. The solutions then depend on whether $b$
vanishes or not:

\begin{itemize}

\item $b=0$. There is a chiral zero mode with $f^R=0$ and a flat
left-handed wave function,
\begin{equation}
f_0^L = \frac{1}{\sqrt{1+\frac{a^L}{2\pi R}}} \, .
\end{equation}
When $c=0$, the wave functions for massive modes
are\footnote{Strictly, there is no uniform convergence as $\eps
\rightarrow 0$ due to the behaviour around $y=0$. We write the
limiting functions for $|y|>\eta$, with $\eta$ an arbitrarily small
positive number, keeping in mind that
the interactions of the wave functions with the brane are to be
calculated before taking the limit $\eps\rightarrow 0$.}
\begin{align}
f_n^L=& B_n[\cos(m_n y) + \tan(m_n \pi R) \sin(m_n |y|)] \, , \\
f_n^R=& C_n[\sin(m_n y) - \sigma(y) \tan(m_n \pi R) \cos(m_n y)] \, ,
\end{align}
with $\sigma$ the sign function and masses given by
\begin{equation}
\tan(\pi R m_n) + \sqrt{\frac{a^L}{a^R}} \tan(\sqrt{a^L a^R} \,
\frac{m_n}{2})=0 \, .
\end{equation}
These solutions reduce to the ones for even scalar fields when
$a^R=0$. Note also that $a^R$ is irrelevant when $a^L=0$. The reason
is that only when $a^L$ forces a discontinuity at the brane is the
``odd-odd'' $a^R$ term nonvanishing for $\eps \rightarrow 0$.
When $c>0$, the solutions reduce again to the ones with $a^R=0$, \ie,
the only effect of a positive $c$ is to cancel the effect of $a^R$.

\item $b>0$. In this case, the wave functions (for $|y|>\eta>0$, see
previous footnote) and masses reduce to the ones without brane terms:
There is a chiral zero mode 
\begin{equation}
f_0^L = 1 \, ,
\end{equation}
and a tower of massive modes
\begin{align}
f_n^L =& B_n \cos(m_n y)  \, , \\
f_n^R =& B_n \sin(m_n y) \, ,
\end{align}
with $m_n=n/R$. In all cases, $f_n(0)=0$, so that the functions are
not piecewise continuous. More precisely, 
\begin{equation}
\lim_{\eps\rightarrow 0} \int \dif y \; \del_\eps(y) f^L_{\eps\,n}(y)
= 0 \, . 
\end{equation}

\end{itemize}

Again, we see that the solutions change abruptly when $b$ or $c$ are
turned on. Moreover, when $b>0$ the wave functions vanish at $y=0$,
so the fermions do not couple to brane fields.


\section{Brane kinetic terms in supersymmetric theories}
\label{susy}

In this section we study the impact of brane kinetic terms in
supersymmetric theories. Besides the intrinsic interest of these
theories, one might hope that supersymmetry improves the
singular behaviour we have 
observed in the previous section. One more reason to
look at supersymmetric models is the following: we have seen that
bosons 
and fermions with brane kinetic terms have different KK wave functions
and masses; an intriguing question is then how supersymmetry
makes the spectra of bosons and fermions in the same supermultiplet
identical. 

We consider the supersymmetric quadratic actions of
massless 5D hyper and vector multiplets including all possible
$\mathcal{O}(1/\Lambda)$ brane kinetic terms.

\subsection{Hypermultiplet}
Off shell, 
the 5D hypermultiplet contains two complex scalars, $\phi_{1,2}$,
a Dirac fermion $\psi$ (with two Weyl components,
$\psi_{1,2}$), and two complex auxiliary 
fields, $F_{1,2}$. We want to
construct an action for these fields, 
invariant under 5D N=1 supersymmetry in the bulk and under 4D N=1
supersymmetry on the brane. It is easy to write such an action in
terms of two 4D 
N=1 superfields~\cite{Arkani-Hamed:2001tb,Marti:2001iw}. The
Lagrangian is 
\begin{equation}
\mathcal{L} = \mathcal{L}_{\mathrm{bulk}} +
\mathcal{L}_{\mathrm{brane}} \, , 
\end{equation}
with
\begin{equation}
\mathcal{L}_{\mathrm{bulk}}=\int d^4 \theta (\Phi^\dagger_1 \Phi_1 +
\Phi^\dagger_2 \Phi_2) + \left[ \int d^2 \theta \Phi_1 \d_y \Phi_2 + 
\hc\right] \, ,
\end{equation}
and
\begin{eqnarray}
\mathcal{L}_{\mathrm{brane}}&=&\delta_0 \left\{ 
\int d^4 \theta \big(a_1 \Phi^\dagger_1 \Phi_1 +
a_2 \Phi^\dagger_2 \Phi_2 \big) \right. \nn
&& ~~ \left. \mbox{} + \left[ \int d^2 \theta \big(b \Phi_1
\d_y 
\Phi_2 - c \Phi_2 \d_y \Phi_1 \big) + \hc \right] \right\} \, ,
\end{eqnarray}
where, in the $y$ basis, $\Phi_i = \phi_i + \sqrt{2} \theta \psi_i +
\theta^2 F_i$.  $\Phi_1$ and $\Phi_2$
(and their corresponding components) have
opposite $Z_2$ parities. In this case we have allowed for complex $b$
and $c$ to keep track of the analytic properties of the supersymmetric
theory. In components, the bulk and brane Lagrangians read (summation
over $i$ is undertood) 
\begin{eqnarray}
&& \mathcal{L}_{\mathrm{bulk}}= -\phi_i^\dagger \Box \phi_i + i \d_\mu
\bar{\psi}_i 
\bar{\sigma}^\mu \psi_i + F_i^\dagger F_i \nn
&& \phantom{\mathcal{L}_{\mathrm{bulk}}=} + \big[ \phi_1 \d_y F_2 +
F_1 \d_y \phi_2 - \psi_1 \d_y \psi_2 + \hc \big] \, , \\
&& \mathcal{L}_{\mathrm{brane}}=\delta_0 \Big\{ a_i
\big[-\phi_i^\dagger \Box 
\phi_i + i \d_\mu \bar{\psi}_i \bar{\sigma}^\mu \psi_i + F_i^\dagger
F_i \big]  \nn
&&  \phantom{\mathcal{L}_{\mathrm{brane}}=} +\big( b\big[\phi_1 \d_y F_2
+ F_1 \d_y \phi_2 - \psi_1 \d_y \psi_2\big] \nn
&& \phantom{\mathcal{L}_{\mathrm{brane}}=} - c \big[\phi_2 \d_y F_1 +
F_2 \d_y 
\phi_1 - \psi_2 \d_y \psi_1 \big] + \hc \big) \Big\} \, .
\end{eqnarray}
The equations of motion for the auxiliary fields are
\begin{eqnarray}
&& [1+a_1 \delta_0] F_1 + [1+b^* \delta_0] \d_y \phi_2^\dagger + c^*
\d_y [\delta_0 \phi_2^\dagger] =0 \, , \\
&& [1+a_2 \delta_0] F_2 - [1+c^* \delta_0] \d_y \phi_1^\dagger - b^*
\d_y [\delta_0 \phi_1^\dagger] =0 \, .
\end{eqnarray}
We can directly solve them, keeping in mind that the delta functions
are regularized. Inserting the solutions in the action we find 
\begin{eqnarray}
&& \mathcal{L} = \left\{- (1+a_1\del_0) \phi^\dagger_1 \Box \phi_1 -
\frac{|1+c\del_0|^2}{1+a_2\del_0} \d_y\phi_1^\dagger \d_y \phi_1
\right. \nn
&& \phantom{\mathcal{L} = \Big\{ } -
\left[\frac{b(1+c^*\del_0)}{1+a_2\del_0} \d_y 
\phi_1^\dagger \d_y(\del_0 \phi_1) + \hc \right]  \nn
&& \left. \phantom{\mathcal{L} = \Big\{ } - \frac{|b|^2}{1+a_2 \del_0}
\d_y(\del_0 
\phi_1^\dagger) \d_y(\del_0 \phi_1) + (1\leftrightarrow 2,
b\leftrightarrow c) \right\} \nn
&& \phantom{\mathcal{L} = \Big\{ } + \bar{\psi}_L i \dsl \psi_L +
\bar{\psi}_R i \dsl \psi_R  
- \bar{\psi}_L \d_y \psi_R + \bar{\psi}_R \d_y \psi_L  \\
\label{susyexact} 
&& \phantom{\mathcal{L} = \Big\{ } + \del_0 \Big[
a_1 \bar{\psi}_L i \dsl \psi_L + a_2 \bar{\psi}_R i \dsl \psi_R
+ \left(- b \d_y \bar{\psi}_R \psi_L + c \bar{\psi}_R \d_y \psi_L + \hc
\right) \Big] \, . \nonumber
\end{eqnarray}
We have defined the chiral 4-component spinors $\psi_L^T=(\psi_1\;0)$,
$\psi_R^T=(0\; \bar{\psi}_2)$. The fermionic sector coincides with
the one studied in Section~\ref{exact} (with $b\rightarrow 2b$,
$c\rightarrow 2c$). That the spectra of $\phi_1$, $\phi_2^\dagger$ and 
$\psi_{L,R}$ are identical can be readily seen from the equations of
motion for the KK wave functions. The equation for $\phi_1$ is
\begin{align}
&\Big[|1+(b+c) \delta_0|^2 
\partial_y^2 + 
\big[(1+(b^\ast+c^\ast)\delta_0)(2b+c) +(1+(b+c)\delta_0)c^\ast
\nonumber \\
&
-\frac{a_2|1+(b+c) \delta_0|^2 }{1+a_2 \delta_0} \big] \delta_0^\prime
\partial_y 
+b(1+(b^\ast+c^\ast) \delta_0)(\delta_0^{\prime \prime}-\frac{a_2
(\delta_0^{\prime})^2}{1+a_2 \delta_0})+b c^\ast (\delta_0^\prime)^2
\Big] f_n^1 
\nonumber \\
&
= -m_n^2 (1+a_1 \delta_0)(1+a_2  \delta_0) f_n^1,
\end{align}
and the one for $\phi_2^\dagger$ can be obtained from this one by
$1\leftrightarrow 2$ and $b \leftrightarrow c^\ast$. For the fermions, 
\begin{align}
\big\{\big[1+(b^\ast+c^\ast)\delta_0 \big]\partial_y+c^\ast
\delta_0^\prime  \big\}f_n^R 
=m_n(1+a_1 \delta_0) f_n^L, \\
\big\{\big[1+(b+c)\delta_0 \big]\partial_y+b \delta_0^\prime
  \big\}f_n^L
=-m_n(1+a_2 \delta_0) f_n^R.
\end{align}
It is then trivial to show that iterating the fermionic equations we
obtain the scalar ones and thus the spectra of fields in the same 4D
N=1 supermultiplet are exactly the
same, as implied by supersymmetry. 
We see that in order for the scalar
solutions to reproduce the fermionic ones, higher order
(singular) terms are required in the scalar sector. They are
reminiscent of the famous $\del(y)^2$ terms
of~\cite{Mirabelli:1997aj}.  

This example illustrates the fact that higher order
operators can modify radically the spectrum, although
supersymmetry does not prevent the singular behaviour we observed in
the last section for the coefficients $b$ and $c$. Nevertheless, we
can consider the theory with $b=0$ and $c=0$, which behaves
smoothly and looks stable under
quantum corrections, as these terms are protected by
nonrenormalization theorems. Note, however, that higher order D-terms,
which can in principle be induced, may reintroduce the singular
behaviour.



\subsection{Vector multiplet}
The 5D off-shell vector multiplet consists of a 5D vector $A_M$, two
Weyl gauginos $\lambda_{1,2}$, a real scalar $\Sigma$, a real
auxiliary field $D$ and a complex one $F_{\chi}$. Under D=4 N=1
supersymmetry they form a vector supermultiplet, $V$, and a chiral one,
$\chi$~\cite{Arkani-Hamed:2001tb,Marti:2001iw}:
\begin{eqnarray}
&& V=-\theta \sigma^\mu\bar{\theta} A_\mu - i \bar{\theta}^2 \theta
\lambda_1 + i \theta^2 \bar{\theta} \bar{\lambda}_1 + \frac{1}{2}
\bar{\theta}^2 \theta^2 D \, , \\
&& \chi = \frac{1}{\sqrt{2}}(\Sigma+iA_5)+\sqrt{2} i \theta \lambda_2 +
\theta^2 F_\chi \, ,
\end{eqnarray}
where $V$ is written in the Wess-Zumino gauge and $\chi$ in the
``$y$'' basis. Since we are interested in the free action we consider
an Abelian theory. The gauge transformations are
\begin{eqnarray}
&& V \rightarrow V+\Lambda+\Lambda^\dagger \, , \\
&& \chi \rightarrow \chi + \sqrt{2} \d_y \Lambda \, ,
\end{eqnarray}
with $\Lambda$ a chiral superfield. The gauge invariant Lagrangian
including $\mathcal{O}(1/\Lambda)$ arbitrary brane terms is
$\mathcal{L} = 
\mathcal{L}_{\mathrm{bulk}} + \mathcal{L}_{\mathrm{brane}}$ with
\begin{equation}
\mathcal{L}_{\mathrm{bulk}} = -\frac{1}{4} \int \dif^2 \theta W^\alpha
W_\alpha + \hc + \int \dif^4 \theta \left(\d_y V -
\frac{1}{\sqrt{2}} (\chi+\chi^\dagger) \right)^2 \, ,
\end{equation}
and
\begin{equation}
\mathcal{L}_{\mathrm{brane}} = \delta_0 \left[  -\frac{1}{4} a \int
\dif^2 
\theta W^\alpha W_\alpha + \hc + c \int \dif^4 \theta \left(\d_y V -
\frac{1}{\sqrt{2}} (\chi+\chi^\dagger) \right)^2 \right] \, .
\end{equation}
The corresponding action is 5D N=1 supersymmetric in the bulk and 4D
N=1 supersymmetric on the brane. $V$ and $\chi$ have opposite $Z_2$
parity. In components the Lagrangian reads
\bea
\mathcal{L}_{\mathrm{bulk}} & =& -\frac{1}{4} F^{MN} F_{MN} +
\frac{1}{2}\d^\mu\Sigma \d_\mu \Sigma + \frac{1}{2} D^2 - \Sigma \d_y
D + F_\chi F_\chi^\dagger \nn
&& \mbox{} - \lambda_1 i \sigma^\mu \d_\mu
\bar{\lambda}_1  - \lambda_2 i \sigma^\mu \d_\mu \bar{\lambda}_2 +
\lambda_2 \d_y \lambda_1 +  \bar{\lambda}_2 \d_y \bar{\lambda}_1 \, ,
\\ 
\mathcal{L}_{\mathrm{brane}} & =& \delta_0 \left[ - \frac{a}{4}
F^{\mu\nu} 
F_{\mu\nu} - \frac{c}{2} F^{\mu 5}F_{\mu 5} +
\frac{c}{2}\d^\mu\Sigma \d_\mu \Sigma + \frac{a}{2} D^2 - c \Sigma \d_y
D + c F_\chi F_\chi^\dagger \right. \nn
&& \left. \mbox{} - a \lambda_1 i \sigma^\mu \d_\mu
\bar{\lambda}_1  - c \lambda_2 i \sigma^\mu \d_\mu \bar{\lambda}_2 +
c \lambda_2 \d_y \lambda_1 + c \bar{\lambda}_2 \d_y \bar{\lambda}_1
\right] \, .
\eea
Using the equations of motion for the auxiliary fields, $F_\chi=0$ and
$(1+a \del_0) D = - \d_y[(1+c\del_0) \Sigma]$, we find the following
Lagrangian for the dynamical fields:
\bea
\mathcal{L} &=& -\frac{1}{4} (1+a\del_0) F^{\mu\nu} F_{\mu\nu} +
\frac{1}{2} 
(1+c\del_0) F^{\mu 5} F_{\mu 5} \nn
&& \mbox{} - \frac{1}{2} (1+c\del_0) \d^\mu \Sigma
\d_\mu \Sigma - \frac{1}{2} \frac{1}{1+a\del_0} \d_y
[(1+c\del_0)\Sigma]\d_y [(1+c\del_0)\Sigma] \\
&& \mbox{} + (1+a \del_0) \bar{\lambda}_L i \dsl \lambda_L +
(1+c\del_0) 
\bar{\lambda}_R i \dsl \lambda_R + (1+c\del_0) (\bar{\lambda}_R \d_y
\lambda_L + \d_y \bar{\lambda}_L \lambda_R) \, , \nonumber
\eea
where we have defined the chiral 4-component spinors
$\lambda_L^T=(\lambda_1,0)$ and $\lambda_R^T=(0,\bar{\lambda}_2)$. 

Again, the equations of motion for bosons and fermions in
the same 4D supermultiplet are identical, which leads to the expected
degeneracy of the spectra. Indeed, the relevant differential equations
are
\begin{equation}
\partial_y \Big[(1+c \delta_0)\partial_y\Big] f_n^A=-m_n^2 (1+a
\delta_0)f_n^A,
\end{equation}
for the gauge boson,
\begin{equation}
\partial_y^2 \Big[(1+c \delta_0) f_n^S \Big]=
\frac{a \delta_0^\prime}{1+a\delta_0} \partial_y \Big[ (1+c \delta_0)
f_n^S \Big] 
-m_n^2 (1+a
\delta_0)f_n^S,
\end{equation}
for the scalar, and
\begin{align}
&\partial_y f_n^L=-m_n f_n^R, \\
& \partial_y \Big[(1+c\delta_0) f_n^R\Big]
=m_n (1+a\delta_0)f_n^L,
\end{align}
for the fermions.
If we now iterate the first order fermionic differential equations we
find that they are identical to the ones of their bosonic
counterparts.

Using the results in the previous section we observe that the KK
masses of all the fields are the same as the ones for gauge bosons. In
particular, they do not exhibit a $b$-like behaviour\footnote{For even
$\lambda_R$ this is so in the fermionic sector because
supersymmetry fixes $a^R=c$.}. This can be understood as the
combination of gauge invariance, N=1 supersymmetry relations and the
fact that both chiralities combine to produce 4D Dirac masses. On
the other hand, there is no finite solution when $c<0$, and the wave
functions 
of even $\Sigma$ and $\lambda_R$ are forced to vanish at $y=0$ when
$c>0$.

To summarize this section, supersymmetry improves the behaviour of the
KK masses and wave functions, but it does not cure it.


\section{Singular solutions, the thin brane limit and perturbation
theory} 
\label{singular}

We have seen that the brane operators which contain derivatives with
respect to $y$ give rise to solutions behaving in a quite singular
way. In this section we study in greater detail a simple example which
exhibits the essential features of the general cases studied above. 

Consider the following free Lagrangian of a 5D massless
fermion: 
\bea
\mathcal{L} &=& \bpsi_L i \dsl \psi_L + \bpsi_R i 
\dsl \psi_R - \bpsi_L \d_y \psi_R + \bpsi_R \d_y \psi_L  \nn
&&  \mbox{} - \frac{b}{2} \delta_0 [\bpsi_L \d_y \psi_R + (\d_y
\bar{\psi}_R) \psi_L] \, . \label{actionexample}
\eea
We take $\psi_L$ even and $\psi_R$ odd. The coefficient of the brane
term, $b$, is naturally of order $1/\Lambda$. Once more, we assume
that the delta function is regularized. The equations of motion can
then be derived in the usual way. They read
\bea
&& \big( 1+\frac{b}{2} \del_0 \big) \d_y \psi_R = i\dsl \psi_L \, ,
\label{eomR} \\ 
&& \big[\big(1+\frac{b}{2} \del_0\big) \d_y + \frac{b}{2}
\del_0^\prime \big] \, \psi_L 
= - i\dsl \psi_R \, .
\eea
These equations admit a zero mode solution:
$\psi_{L,R}(x,y)=f^{L,R}(y) 
\xi_{L,R}(x)$, with $\dsl \xi_{L,R}=0$. This solution is chiral since 
the boundary conditions for 
the odd component, $\psi_R(0)=\psi_R(\pi R)=0$, together with
\refeq{eomR} with a vanishing {\it rhs\/}, imply that $f^R=0$
everywhere. This leaves a homogeneous first order equation for
$f^L$ with no further constraint. Using the following
representation of the delta function,
\begin{equation}
\delta(y) = \frac{\eps}{\pi(y^2+\eps^2)} \, ,
\end{equation}
we readily find
\bea
f^L &=& N \frac{1}{1+\frac{b}{2} \delta_0}  \label{solution} \\
&=& 2 N \pi \frac{y^2+\eps^2}{b \eps + 2 \pi (y^2+\eps^2)}
\label{Pacocorreccion} 
\eea
with $N$ a normalization constant. 
The first equation, \refeq{solution}, holds for any
regularization of the delta function. Observe that when $b<0$ and  
$\eps \leq |b|/2\pi$ in~\refeq{Pacocorreccion}, the function is
singular at the points $y=\pm 
\sqrt{-\eps(b/2 + \pi \eps)/\pi}$; moreover, its integral is
divergent. Therefore we have to restrict $b$ to nonnegative
values. The limit $\eps\rightarrow 0$ then depends only on whether 
$b=0$ or $b>0$. When $b$ vanishes
the limit of \refeq{solution} is simply $f^L(y)=N$. For $b>0$, $f^L$ 
approaches $N$ everywhere except around $y=0$, where it goes to zero
in such a way that $\int \dif y \delta(y) f^L(y)\rightarrow
0$.
Therefore, for positive $b$ the zero mode of the fermion does not
couple to fields living on the brane. Moreover, it is insensitive to
other possible brane terms and to the precise (positive) value of $b$
itself. 

Obviously, in the thin brane limit, the zero mode is not smooth in
the coefficient $b$: the solution for a small positive value of $b$
differs drastically (around the brane)
from the one for vanishing $b$, and there is no finite solution for
any negative $b$. On the other hand, for a thick brane with $\eps >
|b|/2\pi$, the zero mode is well defined and smooth in $b$.

This behaviour also shows up in the limit of the perturbative
expansion in $1/\Lambda$ of the finite $\eps$ result \refeq{solution}:
\begin{equation}
f^L= N \left[1+\sum_{n=1}^\infty \left(-\frac{1}{2}\right)^n
\delta_0^n b^n\right] 
\label{deltaexpansion} \, .
\end{equation}
We see that the breaking of perturbation theory
is caused by the functions
$\delta_0^n$, $n>1$, which are singular when $\eps\rightarrow
0$. Strictly, the expansion~\refeq{deltaexpansion} is only valid for
small $|b|/\eps$, and therefore it is incompatible with the limit
$\eps\rightarrow 0$. However, the situation is analogous to the one in
standard (renormalizable or nonrenormalizable) quantum field theories,
where the quantum UV divergences make the loop expansion badly
defined. Nevertheless, in these theories it is well known that one can
work order by order after renormalization: each order is finite
and (at weak coupling) smaller than the lower ones. Since our
classical divergences have 
essentially the same origin as the usual quantum ones (the unknown UV
of the theory), it is natural to apply the same renormalization
procedure to obtain well defined, finite predictions at each order in
perturbation theory. We develop this point of view in the next
section.


\section{Renormalization and critical theories}
\label{renormalization}

The divergences we have encountered in perturbation theory are
actually a common feature of field theories with infinitely thin
defects. They appear, for example, in classical electrodynamics with
point sources. These singularities indicate a breakdown of the
effective theory at scales in which the details of the fundamental
theory---for instance, in our case, a finite thickness of the brane or
stringy effects---cannot be neglected. In this sense, they are
analogous to the usual UV divergences in radiative
corrections. In~\cite{Goldberger:2001tn}, 
Goldberger and Wise have shown how to deal with classical divergences
in brane theories of codimension greater than one by the usual
renormalization procedure of quantum field theory. The same idea can
be applied to brane theories of codimension one with more singular
operators: adding appropriate counterterms it is possible to cancel
the $\del(0)$ singularities order by order. Let us see how this can be
achieved to $\mathcal{O}(1/\Lambda^2)$ in the example of the previous
section. In this simple case it is sufficient to add to the Lagrangian
the following counterterms: 
\begin{equation}
\mathcal{L}_{\mathrm{ct}} \supset -\frac{b^2}{4} \del_0^2 \bpsi_L \mathrm{i}
\dsl \psi_L 
-\frac{b^2}{4} \del_0^2 \left( \bpsi_L \d_y \psi_R + \hc \right) \,
. \label{counterterms} 
\end{equation}
Then, it is straigthforward to check that, to second order, the zero
mode solution is well defined:
\begin{equation}
f^L = N \big(1-\frac{b}{2} \del_0 + \mathcal{O}(1/\Lambda^3) \big) \, ,
\end{equation}
and the kinetic term in the action is finite:
\begin{equation}
\int_{-\pi R}^{\pi R} \dif y \; (1-\frac{b^2}{4} \del_0^2) \, (f^L)^2 = N^2
\big(2\pi R - b + \mathcal{O}(1/\Lambda^3) \big) \, .
\end{equation}
This example is quite trivial, but it illustrates the main
properties of the counterterms:
\begin{enumerate}

\item They are operators of higher dimension. This is related to the
fact that the brane coefficients are dimensionful and the theory is
nonrenormalizable.

\item The divergences are not pure numbers, but functions of $y$
localized at the branes, \ie, functions like $\del_0^2$. Notice that
also one loop 
corrections in an orbifold without brane terms require the
introduction of counterterms in a nontrivial background: the delta
functions at the fixed points.

\item The coefficients of these singular operators are determined (in
a given regularization) by
the requirement that the divergences must be cancelled. This is
analogous 
to the fact that the infinite parts of usual counterterms are 
fixed by the structure of the radiative corrections. However, in
our case there is no freedom to adjust at will the finite part of the
coefficients, as the operator itself is ``infinite''. This has a very
important consequence:

\item The coefficients of the higher dimensional operators with more
than one delta are not independent parameters of the effective
theory. They do not require new observables to be fixed by
experiment. 

\end{enumerate}
Of course, a full calculation at a given order will include not only
the classical part we have considered so far, but also
loop diagrams and, possibly, contributions from higher order
operators in the effective action. All these contributions may
include new thin brane singularities, which should be cancelled. This
can be achieved adjusting the coefficients of the counterterms
accordingly. In particular, they will contain quantum infinities. In
this way, the usual quantum divergences and the thin brane
singularities are subtracted all at once.

It should be noted that the renormalization procedure we propose is not
identical to the one in~\cite{Goldberger:2001tn}, where the classical
divergences are 
just numbers which can be absorbed into the bare first order
couplings. In fact, the 5D theories 
we are analising differ from the 6D model in~\cite{Goldberger:2001tn}
in at least two 
important aspects. First, we find power-law, instead of logarithmic
divergences. Second, we are dealing with operators with a richer
structure, due to the inclusion of derivatives orthogonal to the
branes. This seems to require in the counterterms nontrivial
backgrounds with more than one delta function.

We would also like to comment on the similarities and differences of
our approach to the one in~\cite{Lewandowski:2001qp}. There, the
singularities in 5D theories with 
brane operators, like the ones we are discussing, were
analised in perturbation theory. The authors proposed to substitute
the singular brane terms by 
terms without $y$ derivatives, giving the same solutions to the
equations of motion outside the core of the brane. The resulting
action was interpreted as a renormalized action. However, we observe
that this ``massage'' (we follow the terminology
of~\cite{Lewandowski:2001qp}) by itself does not lead to finite solutions
since in the original action the divergences near the branes induce
divergences also for the solutions far from the branes, and even for
observables such as KK masses. In the ``massaged'' theory this divergent
behaviour comes from explicit divergences in the coefficients of the
nonsingular operators, which translate into divergencies in the
boundary conditions just outside the core of the branes. On the other
hand, subtracting the explicit divergences in the mentioned
coefficients leads to really finite results. This can be achieved by
adding counterterms such as the ones in~\refeq{counterterms}, which has
the additional advantage that the behaviour inside the brane is also
well defined. We should remark that, in their application of
``massaging'' to the study of the stability of the compact
Randall-Sundrum model, the authors of~\cite{Lewandowski:2001qp}
considered arbitrary {\em 
finite\/} coefficients of operators without $y$ derivatives. Hence,
they did use a renormalized theory.

In Appendix~B we work out the renormalization of a more involved
example: the propagator of a scalar field in the presence of a ``$b$''
brane term. A more complete study of renormalization of
theories with brane terms will be presented
elsewhere.

This renormalization process can be interpreted in a complementary
way. We have seen that, in general, theories with brane kinetic terms
have a singular behaviour in the thin brane limit (before
renormalization), 
such that perturbation theory is spoiled. However, there is
a subclass of ``critical'' theories which are well behaved
and predictive: 
the ones with a tower of operators identical to the
counterterms introduced above (plus all the operators present
initially). The coefficients of these operators are not independent
parameters: both their finite part and the part depending on the
quantum renormalization scale are functions of the coefficients of
bulk operators and brane operators with only one delta. In this sense,
these critical effective theories correspond to a particular {\em
reduction of couplings}~\cite{Zimmermann:1984sx,Oehme:yy}. In fact,
such relations should exist in the low 
energy effective theory if the fundamental theory is to be well
behaved for infinitely thin branes. For instance,
in~\cite{Horava:1996ma} it was hoped that M-theory would directly give
well defined quantities without the $\del(0)$ that were observed in
the supergravity description of the Ho\v{r}ava-Witten
model.\footnote{It 
might be objected that supersymmetric theories cannot be 
critical---or, equivalently, renormalization breaks
supersymmetry---since in our 
discussion above supersymmetry determined completely the higher order
operators of the {\em on-shell\/} theory, and we have seen that they
do not eradicate the singular behaviour. This is actually not so, 
since we included only first order 
operators in the {\em off-shell\/} action. Higher order off-shell
operators 
compatible with supersymmetry can give rise to a critical
theory. Of course, the on-shell supersymmetry transformations will be
different from the ones without the new operators.} A particularly
simple example of critical theory (at least at the classical level),
in which all the coefficients of 
operators of order higher than one vanish, is
the one for a gauge boson with $c=0$. Indeed, in this case one can
directly perform the KK reduction to all orders in $a$, as done
in~\cite{Carena:2002me} and in Section~\ref{exact}.

\section{Conclusions}
\label{conclusions}

We have studied 5D theories for spins $\leq 1$ with general brane
kinetic terms. We have shown that some terms which had been neglected
in the past play an active role, especially in the case of
fermions. We have first calculated the KK masses and wave functions in
these theories to all orders in the coefficients of the brane terms,
using a combination of analytical and numerical methods. 
We have found that some terms change smoothly the solutions without
brane terms, 
while others---the ones with derivatives with 
respect to $y$---change them abruptly, destroying the perturbative
hierarchy of the effective theory. We have also built supersymmetric
free actions with brane terms for the hyper and vector multiplets and 
studied their KK decompositions. Supersymmetry does
not solve in general the difficulties produced by the second kind of
operators, 
although it can alleviate them in some cases. We have then discussed
in detail a 
particular case, showing that the singular behaviour is caused by the
presence of infinitely thin branes. If we insist on performing
perturbative calculations, we have to face $\del(0)$-like divergences
at $\mathcal{O}(1/\Lambda^2)$ and higher orders. At this point, at
least three different paths can be followed. 

First, we can simply take the exact results at their
faith value and accept a theory with such behaviour. From a
phenomenological point of view, the possible realistic models would be
constrained by the decoupling of the wave functions from the branes
and, in the nonsupersymmetric scenario, the absence of scalar zero
modes. In particular, bulk fermions would not couple to a boundary
Higgs, while a nonsupersymmetric bulk Higgs would not have the
required zero mode. Even
more problematic is the fact that perturbation theory breaks
down. It seems very difficult, if not impossible, to construct a
predictive interacting field theory in extra dimensions without some
hierarchy controlling the size of the
operators. 

Second, we can work with thick branes instead of zero width
branes (see~\cite{Kolanovic:2003da} for recent calculations with
kinetic terms on 
thick branes). This is 
natural if the brane is a domain wall. The 
disadvantage of this approach is that the results depend on the
substructure of the brane and are thus very model
dependent. Furthermore, we would like to be able to define sensible
field theories in orbifold compatifications with vanishing brane
width. 

Third, we can renormalize the theory to render it free from both
classical and quantum divergences in perturbation theory. This is
equivalent to working with 
a ``critical'' theory containing a tower of operators whose
coefficients satisfy certain relations. Here we have just argued
that this is a sensible approach to construct extra dimensional models
with thin branes. We have not described the renormalization program in
detail nor proven that it can be carried out to all orders. This
subject is under investigation.

Including mass terms in the bulk and on the branes may be relevant for
phenomenology, but it does not alter in an essential way the results
given here. The same holds for brane kinetic branes on the second
brane. Finally, in a complete model one should also take into account
brane interactions, which may require the introduction of new
counterterms. 

\section*{Acknowledgements}
We thank Ferruccio Feruglio, Luca Girlanda, Ignacio Navarro,
Toni Riotto and Ver\'onica Sanz for discussions.  This work has been
supported in part by MCYT 
under contract FPA2000-1558, by Junta de Andaluc{\'\i}a group FQM 101,
by the European Community's Human Potential Programme under 
contract HPRN-CT-2000-00149 Physics at Colliders, and by PPARC.

\appendix

\section{First order solutions}
\label{firstorderappendix}

In this appendix we present the KK expansions to first order in 
$1/\Lambda$. The first order results are directly finite, but they are
only meaningful when the theory is renormalized, so that higher orders
are also finite.
Perturbatively, the solutions to the generalized eigenvalue problem
can be computed in different 
ways. If the full analytic solutions are known for a smooth
regularization of the brane, we can expand them 
in powers of the dimensionful coefficients $a,b,c$ and then take the 
limit $\epsilon \to 0$. 
Alternatively, we can perform the perturbative expansion
from the very beginning in the eigenvalue equations
and normalization conditions
and solve them order by order (again for a finite $\epsilon$ and
taking the limit of thin brane only at the end of the
calculation). 

We find that the first order KK masses are invariant under
$a \leftrightarrow -b$ for scalars and under $a^L \leftrightarrow -b$
and $a^R \leftrightarrow -c$ for fermions, while the wave functions are
invariant up to possible delta functions.
This can be understood in
the following way. In the scalar sector, one can perform the field
redefinition 
\begin{equation}
\phi \to (1-\frac{b}{2} \delta_0) \phi, \label{scalarfirstorderredef}
\end{equation}
which to first order only has the effect of redefining the
coefficients $b\to 0$ and $a \to a-b$. In the fermion sector, the 
field redefinitions
\begin{eqnarray}
\psi_L &\to& (1-\frac{b}{2} \delta_0) \psi_L,\\
\psi_R &\to& (1-\frac{c}{2} \delta_0) \psi_R,
\end{eqnarray}
redefine $b\to 0$, $a^L\to a^L-b$ and $c\to0$, $a^R\to a^R-c$. All
these field redefinitions generate higher order operators as
well. These are responsible for the singularities observed at higher
orders.

\subsection*{Scalars and Gauge Bosons}
Consider an even scalar with general values of $a,b,c$ (gauge bosons
satisfy the same equations with the condition $b=0$). The solution to
first order is, for the zero mode,
\begin{equation}
f_0(y)=1+\frac{b-a}{4\pi R}-\frac{b}{2} \delta_0,
\end{equation}
and for the massive modes,
\begin{align}
f_n(y)=&\sqrt{2}\cos(m_n^{(0)} y) - \frac{b}{\sqrt{2}} \del_0 \nn
& \mbox{} +\frac{1}{\sqrt{2}}\frac{b-a}{2\pi R} \big[ 
\cos(m_n^{(0)} y)+2m_n^{(0)}(\pi R-|y|) \sin(m_n^{(0)}|y|) \big] ,
\end{align}
with $m_n^{(0)}=n/R$, and masses
\begin{equation}
m_n=\frac{n}{R}\left(1+\frac{b-a}{2\pi R}\right).
\end{equation}
Note that the ``odd-odd'' parameter $c$ has no effect to this order.

For an odd scalar there is no zero mode and the massive ones are
\begin{align}
f_n(y)=\; & \sqrt{2}\sin(m_n^{(0)} y) \nn
& \mbox{} +\frac{1}{\sqrt{2}}\frac{c}{2\pi R} \big[ 
\sin(m_n^{(0)} y)-2m_n^{(0)}\sigma(y)(\pi R-|y|) \cos(m_n^{(0)}y)\big],
\end{align}
with $\sigma(y)$ the sign function and masses
\begin{equation}
m_n = \frac{n}{R}\left(1+\frac{c}{2\pi R} \right).
\end{equation}
The ``odd-odd'' parameters, which are now $a$ and $b$ do not
contribute to this order while the ``even-even'' parameter
$c$ has a non-trivial effect, in contrast with the results in
Section~\ref{exact}. 

\subsection*{Fermions}

For the even component (which we take here to be the LH component),
this case is identical to the scalar case except for the term with
$a^R$, which, just like $c$, does not contribute
to first order. The wave functions and masses to
$\mathcal{O}(1/\Lambda)$ are
\begin{equation}
f_0^L(y)=1+\frac{b-a^L}{4\pi R}-\frac{b}{2} \delta_0,
\end{equation}
for the zero mode (with no zero mode for the odd chirality), and
\begin{align}
f_n^L(y)=\; &\sqrt{2}\cos(m_n^{(0)} y) - \frac{b}{\sqrt{2}} \del_0 \nn
& \mbox{} +\frac{1}{\sqrt{2}}\frac{b-a^L}{2\pi R} \big[ 
\cos(m_n^{(0)} y)+2m_n^{(0)}(\pi R-|y|) \sin(m_n^{(0)}|y|)\big]  , \\
f_n^R(y)=\; &\sqrt{2}\sin(m_n^{(0)} y) \\
& \mbox{} +\frac{1}{\sqrt{2}}\frac{b-a^L}{2\pi R} \big[ 
\sin(m_n^{(0)} y)-2m_n^{(0)}\sigma(y) (\pi R-|y|)
\cos(m_n^{(0)}y)\big], 
\\
m_n =\;&\frac{n}{R}\left(1+\frac{b-a^L}{2\pi R} \right) 
\, ,
\end{align}
for the massive modes.

\section{Renormalization of the scalar propagator}
\label{renappendix}

Here we give another example of classical
renormalization. We consider the scalar theory~\refeq{scalaraction}
with $a=c=0$ for an even scalar, and calculate perturbatively the
classical propagator to $\mathcal{O}(1/\Lambda^2)$, \ie,
$\mathcal{O}(b^2)$. We work in 
momentum space for the first four coordinates and in position space
for the 5th coordinate. The brane delta functions are, as
usual, assumed to be regularized and this time we indicate this
explicitly through a subindex $\eps$, to differentiate these deltas
from the ones which do not need regularization. In perturbation
theory, the propagator can be written 
\begin{equation}
\Delta(p;y_1,y_2)=\Delta_0(p;y_1,y_2)+\Delta_1(p;y_1,y_2) + 
\Delta_2(p;y_1,y_2) + \mathcal{O}(b^3) \, ,
\end{equation}
with $\Delta_n$ proportional to $b^n$.
The zeroth order propagator satisfies the equation
\begin{equation}
[\d_{y_1}^2 + p^2] \Delta_0(p;y_1,y_2) = -\frac{1}{2}
\big[\del(y_1-y_2)+\del(y_1+y_2) \big] \, ,
\label{prop0}  
\end{equation}
with boundary conditions 
\begin{align}
& \left[ \d_{y_1} \Delta_0(p;y_1,y_2) \right]_{y_1=0^+} = 0 \, , ~y_2
\uneq 0 \, , \\
& \left[ \d_{y_1} \Delta_0(p;y_1,y_2) \right]_{y_1=\pi R^-} = 0 \, ,
~y_2 \uneq \pi R \, .  
\end{align}
The solution is
\begin{equation}
\Delta_0(p;y_1,y_2) = - \frac{\cos (p y_<) \cos(p(\pi R-y_>))} {2 p
\sin (p\pi R)} \, ,
\end{equation}
with $y_<=\mbox{Min}\{|y_1|,|y_2|\}$, $y_>=\mbox{Max}\{|y_1|,|y_2|\}$.
The first order propagator is given by the zeroth order propagator
with one insertion of the brane kinetic
term. It reads
\begin{align}
\Delta_1(p;y_1,y_2) &= 
\frac{b}{2} \int \dif z \; \del_\eps(z) \Delta_0(p;y_1,z) \big(
\stackrel{\leftarrow}{\d_z^2} + 
\d_z^2 \big) \Delta_0(p;z,y_2) \nn
&\rightarrow \frac{b}{2} \big[ -2 p^2 \Delta_0(p;y_1,0)
\Delta_0(p;0,y_2) - \del(y_1) \Delta_0(p;y_1,y_2) \nn
&\phantom{\rightarrow \frac{b}{2} \big[} - \Delta_0(p;y_1,y_2)
\del(y_2) \big] \, .
\end{align}
In the second line we have used \refeq{prop0} and taken the limit
$\eps\rightarrow 0$, which is well defined. The singularities arise at
the next order. The second order propagator has two insertions of the
brane kinetic term:
\begin{align}
\Delta_2(p;y_1,y_2) &= 
\frac{b^2}{4} \int \dif z_1 \dif z_2 \; \del_\eps(z_1)  \del_\eps(z_2)
\Delta_0(p;y_1,z_1) \big( \stackrel{\leftarrow}{\d_{z_1}^2} +
\d_{z_1}^2 \big) \Delta_0(p;z_1,z_2) \nn
& \phantom{= \frac{b^2}{4} \int} \times 
\big( \stackrel{\leftarrow}{\d_{z_2}^2} + 
\d_{z_2}^2 \big) \Delta_0(p;z_2,y_2)  \, .
\end{align}
Using \refeq{prop0} for the internal propagator and integrating on
$z_2$, we find a contribution 
\begin{align}
\left[\Delta_2(p;y_1,y_2)\right]_{\mathrm{div}} =&  \frac{b^2}{4} \int
\dif z_1 \; \Big\{ p^2 \del^2_\eps(z_1) \Delta_0(p;y_1,z_1)
\Delta_0(p;z_1,y_2) \nn
& \phantom{\frac{b^2}{4} \int}
+ \d_{z_1}\big[\del_\eps(z_1)
\Delta_0(p;y_1,z_1)\big]\d_{z_1}\big[\del_\eps(z_1)
\Delta_0(p;z_1,y_2)\big] \nn 
& \phantom{\frac{b^2}{4} \int}
-  \del_\eps^2(z_1) \big[\Delta_0(p;y_1,z_1) \big(
\stackrel{\leftarrow}{\d_{z_1}^2} + 
\d_{z_1}^2 \big) 
\Delta_0(p;z_1,y_2) \big] \Big\} \, ,
\end{align}
which diverges when $\eps\rightarrow 0$. To renormalize the
theory to second order (at the classical level) we add the following
counterterms to the Lagrangian:
\begin{equation}
\mathcal{L}_{\mathrm{ct}}= \frac{b^2}{4}\Big\{ - \del_\eps^2(y) \d_\mu
\phi^\dagger \d^\mu \phi - \d_y \big[\del_\eps(y) \phi^\dagger \big]
\d_y \big[\del_\eps(y) \phi \big] + \del_\eps^2(y) \big(\phi^\dagger
\d_y^2 \phi + \d_y^2 \phi^\dagger \phi \big) \Big\} \,
.\label{scalarcounter} 
\end{equation}
Then, the divergences are cancelled and taking $\eps\rightarrow 0$ we
find the finite result
\begin{equation}
\Delta_2^\mathrm{ren}(p;y_1,y_2)=
\frac{b^2}{4} \big[2p^2 \Delta_0(p;y_1,0)+\del(y_1)\big]
\Delta_0(p;0,0) \big[2p^2 \Delta_0(p;0,y_2)+\del(y_2)\big] \, .
\end{equation}
We have checked that the counterterms~\refeq{scalarcounter} also
render the KK reduction finite to order $\mathcal{O}(1/\Lambda^2)$.
The explicit delta functions in the propagator can be translated
through a field redefinition into brane interaction terms. For points
$y_1$ and $y_2$ away from the first brane, the renormalized
classical propagator reads
\begin{align}
\Delta^\mathrm{ren}(p;y_1,y_2) =& \Delta_0(p;y_1,y_2) - b p^2
\Delta_0(p;y_1,0) \Delta_0(p;0,y_2) \nn
& + b^2 p^4 \Delta_0(p;y_1,0) \Delta_0(p;0,0)
\Delta_0(p;0,y_2) + \mathcal{O}(b^3) \nn
=& -\frac{\big[\cos(p y_<) + \frac{b}{2}p \sin(p y_<) \big] \cos \big(p(\pi
R-y_>)\big)} 
{2p\big[\sin(p \pi R)-\frac{b}{2}p \cos(p \pi R) \big]} +
\mathcal{O}(b^3) \, 
. \label{renprop} 
\end{align}
Actually, working out the renormalization to all orders and resumming
the series, it can be shown that  
the exact renormalized propagator is equal to the explicit expression
in the last line of \refeq{renprop}~\cite{inpreparation}. 
The KK masses and wave functions can be easily
obtained from~\refeq{renprop}: they are the poles and the square roots
of the residua of the propagator. They agree with the ones for an even 
scalar when $b=c=0$ and $a=-b$. In fact, the explicit expression in
\refeq{renprop} is identical to the exact propagator of the theory
with $b=c=0$ and $a=-b$. As we have observed in
Appendix~\ref{firstorderappendix}, the field
redefinition~\refeq{scalarfirstorderredef} 
eliminates the $b$ operator from the action~\refeq{scalaraction} and
redefines $a \rightarrow a-b$. It
also introduces new operators at higher orders, which give rise to
singularities. The result in~\refeq{renprop} shows that performing the
field redefinition~\refeq{scalarfirstorderredef} in the
action with the counterterms~\refeq{scalarcounter}, the second order
terms cancel out (at least when $a=c=0$ initially). Conversely,
performing the field redefinition and neglecting the higher order
operators which are generated amounts to a renormalization of the
theory. The delta functions in~\refeq{scalarfirstorderredef} account
for the ones in $\Delta^\mathrm{ren}$.

\end{document}